\documentstyle[12pt,epsfig]{report}

\begin{document}

\begin{titlepage}

\begin{center} {\Large\bf Bound States of a Minimally Interacting Spin-0  and
Spin-1/2 Constituent in the Instantaneous Approximation}

~

G. Bruce Mainland\\ Department of Physics\\
The Ohio State University\\ Columbus, OH 43210, USA
\end{center}

\vspace{.5in}
\centerline{\bf Abstract}

~

Bound-state solutions are obtained numerically in the instantaneous approximation for a spin-0 and
spin-1/2 constituent that interact via minimal electrodynamics. To solve the integral equations in
momentum space, a  method is developed for integrating over the logarithmic singularity in kernels,
making it possible to use basis functions that essentially automatically satisfy the boundary
conditions.  For bound-state solutions that decrease rapidly at small and large values of momentum, 
accurate solutions are obtained with significantly fewer basis functions when the solution is expanded
in terms of these more general basis functions.   The presence of a derivative coupling in single-photon
exchange complicates the construction of the Bethe-Salpeter equation in the instantaneous approximation
and, in the nonrelativistic limit, gives rise to an additional electrostatic potential term that is
second order in the coupling constant and decreases as the square of the distance between constituents.

~

\end{titlepage}

\setcounter{chapter}{1}
\noindent{\Large{\bf 1  Introduction}}

~

There has been a growing interest in solving relativistic, bound-state equations both because important
bound-state systems are relativistic and because the development of high-speed computers makes it
possible to solve such equations.  The Bethe-Salpeter equation [1], which is based on field theory, is
covariant and reduces to the Schr\"odinger equation in the nonrelativistic limit, is the appropriate
equation to use in describing relativistic bound states.  Unfortunately, even numerically the two-body
bound-state equation is exceedingly difficult to solve [2].  For this reason various approximations such
as the Blankenbecler-Sugar approximation [3] or the instantaneous approximation[1,4] are often made that
reduce the covariant equation in four dimensions to an approximately-covariant equation in three
dimensions.  In this article  attention is restricted to the instantaneous approximation although the
general methods developed here can be used in the  implementation of other approximation schemes that
reduce the Bethe-Salpeter equation to a three-dimensional equation. 

The instantaneous approximation, which is the approximation that the binding quanta travel
instantaneously between the bound constituents, was first introduced in the original article by Bethe
and Salpeter [1] and was used in their calculation to demonstrate that the Schr\"odinger equation is the
nonrelativistic limit of the Bethe-Salpeter equation. More in the spirit of this work, Salpeter [4] made
the instantaneous approximation to reduce the Bethe-Salpeter equation to a three-dimensional equation
and calculated corrections to the fine structure of hydrogen-like atoms.  The presence of a derivative
coupling in single-photon exchange complicates the construction of the Bethe-Salpeter equation in the
instantaneous approximation and, in addition to terms that are first order in the coupling constant, 
gives rise to two terms that are second order.  The ``seagull'' interaction yields the same two
second-order interactions, but with different strengths.  In the nonrelativistic limit, the second-order
interaction becomes an electrostatic potential term that decreases as the square of the distance between
the constituents.
 
A major difficulty in solving equations numerically when the instantaneous approximation is made arises
because a logarithmic singularity occurs in the kernel of the integral equations.  Gammel and Menzel [5]
overcome the problem by using a special weighting scheme in the neighborhood of the singularity, and Eyre
and Vary [6] introduce a numerical cutoff and then correct for the effects of the cutoff using
perturbation theory. Later Spence and Vary [7] use B-splines [8] as basis functions and perform all
integrals analytically. Since the B-splines are polynomials, their method is restricted to polynomial
basis functions.  In this article a method is used to integrate over the logarithmic singularity that
allows the use of more general basis functions that essentially automatically satisfy the boundary
conditions and are not necessarily polynomials.  This new method, which is very simple conceptually, is
a  generalization of the method introduced in Ref. 7.  For bound states that decrease rapidly at small
and large values of momentum,  accurate solutions are obtained with significantly fewer basis functions
when these more general basis functions are used.
  
Two physical problems that are of immediate interest are constituent models of quarks
and leptons [9,10] and constituent-quark models of mesons [11]. 
Equations that account for some relativistic effects have had success in describing the properties of
both light and heavy mesons [11].  The fact that there are three families of leptons, much as
there are families of elements or families of hadrons, suggests that the leptons might be composite. The
discovery of neutrino oscillations [12] raises the possibility that neutrinos might also be
composite [13].  If the electron, muon and tau are bound states of a single system, the system is
necessarily relativistic:  The  mass of the tau must, of course, be less than the sum of the masses of
the bound, constituent particles.  Since the ratio of the electron's mass  to that of the tau's equals
1/3536,  the  mass of the electron is less than $1/3536$ of the sum of the constituent masses,
indicating highly relativistic binding. 

The Bethe-Salpeter equation discussed here has been solved exactly in the strong binding (zero energy)
limit when the ladder approximation is made [10].  In the instantaneous approximation the numerical
solutions obtained here are not in good agreement with the exact zero-energy solutions so, not
surprisingly, the instantaneous approximation is not satisfactory for very strongly bound states.

To estimate the accuracy of each solution, in the physical region the left- and right-hand sides of the
equation are calculated midway between each knot, and a reliability coefficient $R$ [14], which is a
statistical measure of how accurately the left- and right-hand sides agree at the selected points, is
calculated.  Examining points where the left- and right-hand sides of the equation agree least well
reveals possible problems with solutions and suggests possible remedies.

~

\setcounter{chapter}{2}
\noindent\begin{Large}{\bf 2  Derivation and Separation of the Bethe-- \\
\makebox[.6cm]\  Salpeter Equation in the Instantaneous\\ 
\makebox[.6cm]\ Approximation}
\end{Large}

~

When a spin-0 field $\phi(x)$, which represents a quanta with charge $Q$ and mass $M$,
interacts via minimal electrodynamics with a spin-$1/2$ field $\Psi(x)$,
which represents a quanta with charge $q$ and mass $m$, the renormalizable  Lagrangian
is [15] 
\begin{eqnarray*}
L = : [( \imath \partial^{\mu} - QA^{\mu})\phi ][(-\imath
\partial_{\mu} - QA_{\mu})\phi^+] -M^2\phi^+\phi 
\end{eqnarray*}
\vspace{-0.5cm}
\begin{equation}
\hspace{1.3cm} + \bar{\Psi}\gamma_{\mu}(\imath
\partial^{\mu} - qA^{\mu})\Psi - m \bar{\Psi} \Psi - {1 \over 4}F_{\mu \nu}F^{\mu
\nu} : \hspace{1.6cm}
\end{equation}
\noindent where $F_{\mu \nu} = \partial_\nu A_\mu - \partial_\mu A_\nu$.

~

The two-particle, Bethe-Salpeter wave function is defined by

\setcounter{equation}{1}
\begin{equation}
\chi_K (x_1,x_2) = <0|T(\Psi(x_1)\phi(x_2))|K>.
\end{equation}

\noindent In (2.2) the symbol $T$ represents time ordering and the letter $K$
labels the four-momentum of the bound state.  The center-of-mass coordinates $X^\mu$
are defined by

\begin{equation}
X^\mu = \xi x{^\mu _1} + (1-\xi)x{^\mu _2},
\end{equation}

\noindent and the relative coordinates x$^\mu$ by 

\begin{equation}
x^\mu = x{^\mu _1} - x{^\mu _2}.
\end{equation}

\noindent When the parameter $\xi$ is given by $\xi = m/(m+M)$ , the usual
nonrelativistic definition of center-of-mass coordinates results.  As will be seen,
the parameter $\xi$ drops out of the Bethe-Salpeter equation when the instantaneous
approximation is made so there is no need to make a specific choice.  The dependence
of $\chi_K(x_1,x_2)$ on the center-of-mass coordinates factors with the result that
$\chi_K(x_1,x_2)$ can be rewritten as

\begin{equation}
\chi_K(x_1 ,x_2) = (2\pi)^{-3/2}e^{-iX^\mu K_\mu}\chi_K(x). 
\end{equation}

Denoting the Fourier transform of $\chi_K(x)$ by $\chi_K(p)$, the Bethe-Salpeter
equation is

\setcounter{equation}{5}
\begin{eqnarray*}
(p^\mu\gamma_\mu + \xi K^\mu\gamma _\mu - m)\{[p^\mu - (1-\xi)K^\mu][p_\mu -
(1-\xi)K_\mu]-M^2\}\chi_K(p)\\
\end{eqnarray*}
\vspace{-1.3cm}
\begin{eqnarray*}
\hspace{-1.6cm}={iqQ \over (2\pi)^4}  \int_{- \infty}^{\infty} {d^4q \over
(p-q)^2+i\epsilon}[p^\mu \gamma _\mu + q^\mu \gamma_\mu -
2(1-\xi)K^\mu\gamma_\mu]\chi_K(q)\\
\end{eqnarray*}
\vspace{-1.0cm}
\begin{eqnarray*}
\hspace{-3.6cm}+{4(qQ)^2 \over (2\pi)^8} \int_{-
\infty}^{\infty} {d^4q  \over q^2 -m^2 +i\epsilon} \hspace{0.1in}  
{2m-q^\mu\gamma_\mu
\over (p-q+\xi K)^2+i\epsilon} \times
\end{eqnarray*}
\begin{equation}
\hspace{3.0cm}\int_{-
\infty}^{\infty}{d^4k \over (q-k-\xi K)^2 +
i\epsilon}\chi_K(k).\\
\end{equation}

\noindent A charged, spin-0 boson interacts electromagnetically through two fundamentally different
processes: single-photon exchange and the ``seagull'' interaction.  In the above equation the terms
proportional to $qQ$ and $(qQ)^2$ arise, respectively, from these two interactions. Although the
``seagull'' interaction is second order in the coupling constant, it has been included in the
Bethe-Salpeter equation to determine its effect on solutions.  The spirit of this calculation, then, is
similar to others dealing with the Bethe-Salpeter equation whereby  the ladder approximation is made,
but some solutions with large coupling constants are studied.  The Feynman diagrams for these
two interactions are shown in Fig. 2.1

\begin{figure}
\begin{center}
\epsfig{file=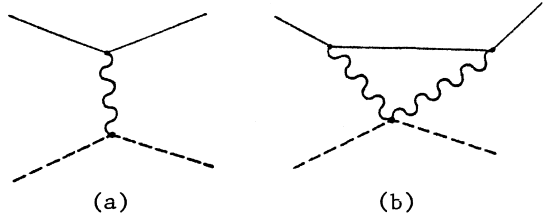}
\end{center}
\vspace{-.5cm}
\caption{Feynman diagrams for (a) single-photon exchange and (b) the ``seagull'' interaction.  The
solid, dashed and wavy lines represent a spin-$1/2$ fermion, a spin-0 boson and a photon, respectively.
\label{figure}} 
\end{figure}

The instantaneous approximation [1,4] is made by making the replacement
\setcounter{equation}{6}
\begin{equation}
{1 \over k^2 + i\epsilon} = {1 \over k{^2_0} - {{\bf k}^2}+i\epsilon} \longrightarrow
{1
\over -{{\bf k}^2}+i\epsilon}
\end{equation}
\noindent in each of the three photon propagators in (2.6).  Defining
\begin{eqnarray*}
\hspace{4.0cm}\Psi_K({\bf k}) \equiv
\int_{-\infty}^{\infty}dk_0\hspace{.1cm}\chi_ K(k),\hspace{4.0cm} (2.8a)
\end{eqnarray*}
and
\begin{eqnarray*}
\hspace{4.0cm}\phi_K({\bf k}) \equiv
\int_{-\infty}^{\infty}dk_0\hspace{.1cm}k_0\hspace{.1cm}\chi_K(k),
\hspace{3.6cm} (2.8b)
\end{eqnarray*}
\noindent the integral over $q_0$ in the single-photon-exchange term in (2.6) and the
integral over $k_0$ in the ``seagull'' term can be  carried out immediately.  In
the ``seagull'' term it is then possible to integrate over $q_0$.  Going to the
center-of-mass frame where $K^\mu = (E,0)$, (2.6) becomes
\begin{eqnarray*}
\hspace{-2.0cm}[p^0\gamma^0 -p^i\gamma^i+ \xi E \gamma _0 - m]\{[p^0 -
(1-\xi)E]^2-p^ip^i-M^2\}\chi_E(p)\\
\end{eqnarray*}
\vspace{-1.3cm}
\begin{eqnarray*}
\hspace{-1.5cm}= - {iqQ \over (2\pi)^4} \int_{- \infty}^{\infty} {d^3q \over ({\bf
p}-{\bf q})^2}}{[\gamma^0p^0 -\gamma^i(p^i+q^i)-2(1-\xi) E\gamma^0]\Psi_E({\bf q})
\\
\end{eqnarray*}
\vspace{-1.3cm}
\begin{eqnarray*}
\hspace{-3.0cm}- {iqQ \over (2\pi)^4} \int_{- \infty}^{\infty}{d^3q \over ({\bf
p}-{\bf q})^2}\gamma^0\phi_E({\bf q})\hspace{4.5cm}\\
\end{eqnarray*}
\vspace{-1.3cm}
\begin{eqnarray*}
\hspace{0.1cm}-{2i(qQ)^2 \over (2\pi)^7}\int_{- \infty}^{\infty}{d^3q \over ({\bf
p}-{\bf q})^2}\hspace{.2cm}{2m + \gamma^iq^i \over \omega_m(q)}
\int_{- \infty}^{\infty}
{d^3k \over ({\bf
q}-{\bf k})^2}\hspace{.2cm}\Psi_E({\bf k}).\hspace{2.3cm} (2.9)\\ 
\end{eqnarray*}

\noindent In the above equation $\chi_E(p)$ is the value of $\chi_K(p)$ in the
center-of-mass frame, etc., and $\omega_m(q) \equiv (m^2+q^2)^{1/2}$.

~

Solving (2.9) for $\chi_E(p)$, integrating over $p^0$ on both sides of the equation
and using (2.8) yields

\begin{eqnarray*}
 {\omega_M(p)E\Psi_E({\bf p})}
= {\omega_M(p) \over \omega_m(p)} [\omega_M (p) + \omega_m (p)] [\gamma^0
\gamma^ip^i + \gamma^0m] \Psi_E({\bf p})\hspace{2.cm}\\
\end{eqnarray*}
\vspace{-1.3cm}
\begin{eqnarray*}
\hspace{-1.5cm}+{qQ \over 16\pi^3} \int_{- \infty}^{\infty}{d^3q \over ({\bf
p}-{\bf q})^2}\{(1-\xi)E+[{\omega_M(p) + \omega_m(p) \over
\omega_m(p)}]\gamma^0\gamma^ip^i\\
\end{eqnarray*}
\vspace{-1.1cm}
\begin{eqnarray*} 
\hspace{1.8cm}+{\omega_M(p) \over \omega_m(p)}\gamma^0m +
\gamma^0\gamma^iq^i\}
\Psi_E({\bf q})\\
\end{eqnarray*}
\vspace{-1.3cm}
\begin{eqnarray*}
\hspace{-1.3cm}-{Qq \over 16\pi^3}\int_{- \infty}^{\infty}{d^3q \over ({\bf
p}-{\bf q})^2} \phi_E({\bf q})\hspace{5cm}\\
\end{eqnarray*}
\vspace{-1.3cm}
\begin{eqnarray*}
\hspace{1cm}-{(qQ)^2 \over (2\pi)^6}\int_{- \infty}^{\infty}{d^3q \over ({\bf
p}-{\bf q})^2}\hspace{.2cm}{2m\gamma^0 + \gamma^0\gamma^i q^i \over \omega_m (q)}
\int_{- \infty}^{\infty}
{d^3k \over ({\bf
q}-{\bf k})^2}\hspace{.2cm}\Psi_E({\bf k}).\hspace{.7cm}(2.10) 
\end{eqnarray*}

~

\noindent Eq. (2.10) cannot readily be  solved  because of the presence of the two
functions $\Psi_E$ and $\phi_E$ that are related as indicated in (2.8).  The
function $\phi_E$ is present, of course, as a consequence of the derivative
coupling. 

~

It is possible, however, to express $\phi_E$ in terms of $\Psi_E$ as follows:  By
dividing both sides of (2.9) by \{$[p^0 -(1-\xi)E]^2-p^ip^i-M^2$\} and
then integrating over $p^0$, a second equation is obtained that involves both
$\Psi_E$ and $\phi_E$: 
\begin{eqnarray*} 
\hspace{-1.5cm}\omega_M(p) \phi_E({\bf p})
= \omega_M(p) \hspace{.2cm}(-\xi E + \gamma^0\gamma^ip^i +
\gamma^0m)\hspace{.2cm}\Psi_E({\bf p})\hspace{1.5cm}
\end{eqnarray*}
\vspace{-.5cm}
\begin{eqnarray*}
\hspace{-1.5cm}+{qQ \over 16\pi^3}\int_{- \infty}^{\infty}{d^3q \over ({\bf
p}-{\bf q})^2}\hspace{.2cm}[(1-\xi)E + \gamma^0\gamma^i(p^i + q^i)]\Psi_E({\bf
q})
\end{eqnarray*}
\vspace{-.3cm}
\begin{eqnarray*}
\hspace{-4.5cm}-{qQ \over 16\pi^3}\int_{- \infty}^{\infty}{d^3q \over ({\bf
p}-{\bf q})^2}\hspace{.2cm} \phi_E({\bf q})\hspace{1.5cm}
\end{eqnarray*}
\vspace{-.4cm}
\begin{eqnarray*}
\hspace{1.cm}-{(qQ)^2 \over (2\pi)^6}\int_{- \infty}^{\infty}{d^3q \over ({\bf
p}-{\bf q})^2}\hspace{.2cm}{2m\gamma^0 + \gamma^0\gamma^iq^i \over \omega_m(q)}
\int_{- \infty}^{\infty}{d^3k \over ({\bf
q}-{\bf k})^2}\hspace{.2cm}\Psi_E({\bf k})\hspace{.8cm}(2.11)
\end{eqnarray*}
\noindent Subtracting (2.11) from (2.10) and solving for $\phi_E({\bf p})$ yields
\begin{eqnarray*}
\hspace{-8.0cm}\phi_E({\bf p}) = (1-\xi)E\Psi_E({\bf p})\hspace{2.1cm}\\
\end{eqnarray*}
\vspace{-1.3cm}
\begin{eqnarray*}
- {1\over
\omega_m(p)}\hspace{.2cm}(\gamma^0\gamma^ip^i + \gamma^0m)
[\omega_M(p)\Psi_E({\bf p})
+{qQ \over 16\pi^3}\int_{-
\infty}^{\infty} {d^3q \over ({\bf
p}-{\bf q})^2}\Psi_E({\bf q})].\hspace{0.3cm}(2.12)\\
\end{eqnarray*}
\noindent Using (2.12) to express $\phi_E({\bf p})$ in terms of $\Psi_E$, (2.10)
becomes the Bethe-Salpeter equation in the instantaneous approximation:
\begin{eqnarray*}
\omega_M(p)E \Psi_E ({\bf p})
={\omega_M(p) \over \omega_m(p)}
\left [ \omega_M (p) + \omega_m (p) \right ] \left [ \gamma^0 \gamma^i
p^i + \gamma^0 m \right ]  E \Psi_E({\bf p})\\
\end{eqnarray*}
\vspace{-1.2cm}
\begin{eqnarray*}
+{qQ \over 16 \pi^3} \int^\infty_{- \infty} {d^3 q \over ( {\bf p} - {\bf
q})^2} ~ \left \{ \left [ { \omega_M (p) \over \omega_m (p)} + 1 \right]
\gamma^0 \gamma^i p^i + \left [ { \omega_M (q) \over \omega_m (q)} +
1 \right ] \gamma^0  \gamma^i q^i \right .\\
\end{eqnarray*}
\vspace{-1.2cm}
\begin{eqnarray*}
\hspace{2.0cm}+ \left . \left [ {\omega_M (p) \over \omega_m (p)} +
{\omega_M (q) \over \omega_m (q)} \right ] \gamma^0m \right \} \Psi_E
(\bf q)\\
\end{eqnarray*}
\vspace{-1.1cm}
\begin{eqnarray*}
\hspace{-0.9cm}+{(qQ)^2 \over 4 (2 \pi)^6} ~ \int^\infty_{-\infty}  ~ {d^3 q \over (
{\bf p} - {\bf q})^2} ~
~ {m \gamma^0 + \gamma^0 \gamma^i q^i \over \omega_m (q)} ~
\int^\infty_{- \infty} ~ {d^3 k \over ( \bf q - \bf k)^2} \Psi_E (\bf
k)\\
\end{eqnarray*}
\vspace{-1.2cm}
\begin{eqnarray*} 
\hspace{0.5cm}- { (qQ)^2 \over (2 \pi)^6} ~ \int^\infty_{- \infty} ~ {d^3 q
\over ( {\bf p} - {\bf q})^2} ~
~ {2m \gamma^0 + \gamma^0 \gamma^i q^i \over \omega_m (q)} ~
\int^\infty_{- \infty} ~ {d^3 k \over ( {\bf q} - {\bf k})^2} \Psi_E ({\bf
k})\hspace{0.9cm}(2.13)\\
\end{eqnarray*}
\noindent In (2.13) the final two terms, which are proportional to
$(qQ)^2$, arise from the derivative coupling in single photon exchange and
the``seagull'' interaction, respectively.  
 
Eq. (2.13) is much easier to solve numerically than the Bethe-Salpeter
equation, primarily because it is much easier to obtain solutions with real energy
eigenvalues.  Specifically, equations of the form (2.13) can be solved
numerically by converting them to matrix eigenvalue equations.  When each side is
multiplied by $\Psi^\dagger_E(\bf p)$ and integrated over $d^3p$, excluding the
eigenvalue $E$, the quantity on the left-hand side is Hermitian and positive definite
and the quantity on the right-hand side is Hermitian.  As a consequence the energy
eigenvalue must be real [16].  The Hermiticity results from the fact that the momenta $\bf p$ and
$\bf q$ appear symmetrically on the right-hand side of (2.13).  In the very special
cases where the Bethe-Salpeter equation possesses the Hermiticity properties of
(2.13), the equation is relatively easy to solve  numerically [2,17] in spite of the
fact that after separation of the two angular variables, it is still an integral or partial
differential equation in two variables.

Solutions to (2.13) are of the form
\begin{eqnarray*}
\hspace{3.3cm}\Psi_E({\bf p}) & = &\left [ \begin{array}{c}
G^{(\pm)} (p) \phi^{(\pm)}(\theta,\varphi)\\[10pt]
F^{(\pm)} (p)\phi^{(\mp)}(\theta,\varphi)\end{array}
\right],\hspace{3.3cm}(2.14)\\ 
\end{eqnarray*}
where the $\phi^{(\pm)}(\theta,\varphi)$ are the same functions [15] that
represent the angular dependence of the bound-state solutions to the
Dirac equation when the potential is spherically symmetric.  After the angular
integration is performed using Hecke's theorem [18,19,20,10], the angular variables
separate.  Multiplying the resulting upper and lower equations by $p$ and $-p$,
respectively, yields
\begin{eqnarray*}
\hspace{-6.5cm}E  \omega_M ({p}) &\left [ \begin{array}{c}
p G^{(\pm)} (p)\\[10pt]
p F^{(\pm)} (p) \end{array} \right]\hspace{7.5cm}\\
\end{eqnarray*}
\vspace{-1.0cm}
\begin{eqnarray*}
 \hspace{-.9cm}=
 {\omega_M (p) \over \omega_m (p)}
[ \omega_M (p) + \omega_m (p) ]
\left \{ p \left [ \begin{array}{c}
p F^{(\pm)} (p)\\
p G^{(\pm)} (p) \end{array} \right]
+ m
\left [ \begin{array}{c}
p G^{(\pm)} (p)\\
- p F^{(\pm)} (p) \end{array} \right] \right \} \hspace{.9cm}
\end{eqnarray*}
\vspace{-1.3cm}
\begin{eqnarray*}
\\[10pt]
+{qQ \over 8 \pi^2} p
\left [ {\omega_M (p) \over \omega_m (p)} +1 \right ] \left [ \begin{array}{c}
\int dq ~
q F^{(\pm)} (q) Q_{j \pm {1\over 2}} \left ( {p^2 + q^2 \over 2pq} \right)\\[8pt]
\int dq ~
q G^{(\pm)} (q) Q_{j \mp {1\over 2}} \left ( {p^2 + q^2 \over 2pq}
\right)\end{array} \right]\hspace{4.0cm}\\[10pt]
\end{eqnarray*}
\vspace{-1.3cm}
\begin{eqnarray*}
\hspace{-1.6cm}+{qQ \over 8 \pi^2} \int q ~ dq
\left [ {\omega_M (q) \over \omega_m (q)} +1 \right ] ~
\left [ \begin{array}{c}
q F^{(\pm)} (q) Q_{j \mp {1\over 2}} \left ( {p^2 + q^2 \over 2pq} \right)\\[8pt]
q G^{(\pm)} (q) Q_{j \pm {1\over 2}} \left ( {p^2 + q^2 \over 2pq} \right)\end{array}
\right]\hspace{1.7cm}\\
\end{eqnarray*}
\vspace{-.8cm}
\begin{eqnarray*}
\hspace{-.9cm}+{qQ \over 8 \pi^2} m~ \int  ~ dq
\left [ {\omega_M (p) \over \omega_m (p)} +
{\omega_M (q) \over \omega_m (q)}
 \right ]
\left [ \begin{array}{c}
q G^{(\pm)} (q) Q_{j \mp {1\over 2}} \left ( {p^2 + q^2 \over 2pq} \right)\\
- q F^{(\pm)} (q) Q_{j \pm {1\over 2}} \left ( {p^2 + q^2 \over 2pq}
\right)\end{array} \right]\hspace{.9cm}\\
\end{eqnarray*}
\vspace{-.7cm}
\begin{eqnarray*}
+{q^2 Q^2 \over 4 (2 \pi )^4} ~ \int {dq \over \omega_m (q)} ~ \int dk
\left \{ m
\left [ \begin{array}{c}
kG^{(\pm)} (k) Q_{j \mp {1\over 2}} \left ( {k^2 + q^2 \over 2kq} \right )
Q_{j \mp{1\over 2}} \left ( {p^2 + q^2 \over 2pq} \right ) \\
-kF^{(\pm)} (k) Q_{j \pm {1\over 2}} \left ( {k^2 + q^2 \over 2kq} \right )
Q_{j \pm {1\over 2}} \left ( {p^2 + q^2 \over 2pq} \right )
\end{array} \right ]\right.\\
\end{eqnarray*}
\vspace{-.8cm}
\begin{eqnarray*}
\hspace{2.4cm} +q
 \left . \left [ \begin{array}{c}
kF^{(\pm)} (k) Q_{j \pm {1\over 2}} \left ( {k^2 + q^2 \over 2kq} \right )
Q_{j \mp {1\over 2}} \left ( {p^2 + q^2 \over 2pq} \right ) \\
kG^{(\pm)} (k) Q_{j \mp {1\over 2}} \left ( {k^2 + q^2 \over 2kq} \right )
Q_{j \pm {1\over 2}} \left ( {p^2 + q^2 \over 2pq} \right )
\end{array} \right ]
\right \}\\
\end{eqnarray*}
\vspace{-.8cm}
\begin{eqnarray*}
-{q^2 Q^2 \over (2 \pi )^4} ~ \int {dq \over \omega_m (q)} ~ \int dk
\left \{2 m \left [ \begin{array}{c}
kG^{(\pm)} (k) Q_{j \mp {1\over 2}} \left ( {k^2 + q^2 \over 2kq} \right )
Q_{j \mp {1\over 2}} \left ( {p^2 + q^2 \over 2pq} \right ) \\
-kF^{(\pm)} (k) Q_{j \pm {1\over 2}} \left ( {k^2 + q^2 \over 2kq} \right )
Q_{j \pm {1\over 2}} \left ( {p^2 + q^2 \over 2pq} \right )
\end{array} \right ]\right.\\
\end{eqnarray*}
\vspace{-.8cm}
\begin{eqnarray*}
\hspace{3.8cm}+ q
 \left . \left [ \begin{array}{c}
kF^{(\pm)} (k) Q_{j \pm {1\over 2}} \left ( {k^2 + q^2 \over 2kq} \right )
Q_{j \mp {1\over 2}} \left ( {p^2 + q^2 \over 2pq} \right ) \\
kG^{(\pm)} (k) Q_{j \mp {1\over 2}} \left ( {k^2 + q^2 \over 2kq} \right )
Q_{j \pm {1\over 2}} \left ( {p^2 + q^2 \over 2pq} \right )
\end{array} \right ] \right \},\hspace{.4cm} (2.15)
\end{eqnarray*}
\noindent where $Q_{j \pm {1\over 2}}$ is a Legendre function of the second kind and
\setcounter{equation}{15}
\begin{equation}
\omega_M(p)\equiv\sqrt{p^2+M^2}, \hspace{.5in}\omega_m(p)\equiv\sqrt{p^2+m^2}. 
\end{equation}
To rewrite (2.15) in terms of dimensionless variables, the two masses are first
rewritten as follows:
\begin{equation} m \equiv m_0 (1 - \Delta), \hspace{.5in} M \equiv m_0 (1 + \Delta).
\end{equation}
\noindent The dimensionless momentum $p'$ is defined by
\begin{equation}
p'\equiv{p \over m_0}, 
\end{equation}
\noindent and the dimensionless energy $\epsilon$ by
\begin{equation}
\epsilon \equiv {E \over M + m} = {E \over 2m_0}.
\end{equation}
\noindent Defining
\begin{eqnarray*}
\hspace{2.95cm}\omega_+(p')\equiv{\omega_M(p) \over
m_0}=\sqrt{(1+\Delta)^2+p'^2},\hspace{2.95cm} (2.20a)
\end{eqnarray*}
\noindent and
\begin{eqnarray*}
\hspace{2.95cm}\omega_-(p')\equiv{\omega_m(p) \over
m_0}=\sqrt{(1-\Delta)^2+p'^2},\hspace{2.95cm} (2.20b)
\end{eqnarray*}
\noindent and omitting primes since all variables are now dimensionless, (2.15) becomes
\begin{eqnarray*}
\hspace{-3.0cm}2\epsilon \hspace{0.05cm} \omega_+ ({p}) &\left [ \begin{array}{c}
p G^{(\pm)} (p)\\[10pt]
p F^{(\pm)} (p) \end{array} \right]\hspace{8.5cm}\\
\end{eqnarray*}
\vspace{-1.0cm}
\begin{eqnarray*}
\hspace{-1.0cm}= {\omega_+ ( p) \over \omega_- ( p)}
[ \omega_+ (p) + \omega_- (p) ]
\left \{ p \left [ \begin{array}{c}
p F^{(\pm)} (p)\\
p G^{(\pm)} (p) \end{array} \right]
+ (1 - \Delta )
\left [ \begin{array}{c}
p G^{(\pm)} (p)\\
- p F^{(\pm)} (p) \end{array} \right]
\right \}\hspace{1.0cm}\\
\end{eqnarray*}
\vspace{-0.8cm}
\begin{eqnarray*}
\hspace{-2.0cm}+{q Q \over 8 \pi^2} \hspace{0.05cm}p
\left [ {\omega_+ (p) \over \omega_- ( p)} +1 \right ] \int d  q
\left [ \begin{array}{c}
q F^{(\pm)} ( q) Q_{j \pm {1 \over 2}} \left ( {p^2 + q^2 \over 2pq} \right)\\[8pt]
q G^{(\pm)} ( q) Q_{j \mp {1 \over 2}} \left ( {p^2 + q^2 \over 2pq}
\right)\end{array} \right]\hspace{3.0cm}\\
\end{eqnarray*}
\vspace{-0.8cm}
\begin{eqnarray*}
\hspace{-1.8cm}+{q Q \over 8 \pi^2} \int q ~ dq
\left [ {\omega_+ (q) \over \omega_- (q)} +1 \right ] ~
\left [ \begin{array}{c}
q F^{(\pm)} (q) Q_{j \mp {1 \over 2}} \left ( {p^2 + q^2 \over 2pq} \right)\\[8pt]
q G^{(\pm)} (q) Q_{j \pm {1 \over 2}} \left ( {p^2 + q^2 \over 2pq}
\right)\end{array}
\right]\hspace{3.0cm}\\
\end{eqnarray*}
\vspace{-0.8cm}
\begin{eqnarray*}
\hspace{-1.3cm}+{q Q \over 8 \pi^2} (1 - \Delta) \int  dq
\left [ {\omega_+ (p) \over \omega_- (p)} +
{\omega_+ (q) \over \omega_- (q)}
 \right ]
\left [ \begin{array}{c}
q G^{(\pm)} (q) Q_{j \mp {1 \over 2}} \left( {p^2 + q^2 \over 2pq} \right)\\[8pt]
- q F^{(\pm)} (q) Q_{j \pm {1 \over 2}} \left( {p^2 + q^2 \over 2pq}
\right)\end{array} \right]\hspace{1.0cm}\\
\end{eqnarray*}
\vspace{-0.8cm}
\begin{eqnarray*}
\hspace{-0.8cm}+{(qQ)^2 \over 4 (2 \pi )^4} ~ \int {dq \over \omega_- (q)} ~ \int dk
 \} (1-\Delta)
\left [ \begin{array}{c}
kG^{(\pm)} (k) Q_{j \mp {1 \over 2}} \left ( {k^2 + q^2 \over 2kq} \right )
Q_{j \mp {1 \over 2}} \left ( {p^2 + q^2 \over 2pq} \right ) \\
-kF^{(\pm)} (k) Q_{j \pm {1 \over 2}} \left ( {k^2 + q^2 \over 2kq} \right )
Q_{j \pm {1 \over 2}} \left ( {p^2 + q^2 \over 2pq} \right )
\end{array}\right]
\end{eqnarray*}
\vspace{-0.2cm}
\begin{eqnarray*}
\hspace{5.0cm}+ q\left.\left[
\begin{array}{c} kF^{(\pm)} (k) Q_{j \pm {1 \over 2}} \left ( {k^2 + q^2 \over 2kq}
\right ) Q_{j \mp {1 \over 2}} \left ( {p^2 + q^2 \over 2pq} \right ) \\
kG^{(\pm)} (k) Q_{j \mp {1 \over 2}} \left ( {k^2 + q^2 \over 2kq} \right )
Q_{j \pm {1 \over 2}} \left ( {p^2 + q^2 \over 2pq} \right )
\end{array} \right ]
\right \}\\
\end{eqnarray*}
\vspace{-0.8cm}
\begin{eqnarray*}
\hspace{-0.7cm}-{(qQ)^2 \over {(2 \pi )^4}} ~ \int {dq \over \omega_- (q)} ~
\int dk
\left \{2 (1-\Delta)
\left [ \begin{array}{c}
kG^{(\pm)} (k) Q_{j \mp {1 \over 2}} \left ( {k^2 + q^2 \over 2kq} \right )
Q_{j \mp {1 \over 2}} \left ( {p^2 + q^2 \over 2pq} \right ) \\
-kF^{(\pm)} (k) Q_{j \pm {1 \over 2}} \left ( {k^2 + q^2 \over 2kq} \right )
Q_{j \pm {1 \over 2}} \left ( {p^2 + q^2 \over 2pq} \right )
\end{array} \right ]\right.\\
\end{eqnarray*}
\vspace{-0.8cm}
\setcounter{equation}{20}
\begin{equation}
\hspace{5.0cm}+ q
 \left . \left [ \begin{array}{c}
kF^{(\pm)} (k) Q_{j \pm {1 \over 2}} \left ( {k^2 + q^2 \over 2kq} \right )
Q_{j \mp {1 \over 2}} \left ( {p^2 + q^2 \over 2pq} \right ) \\
kG^{(\pm)} (k) Q_{j \mp {1 \over 2}} \left ( {k^2 + q^2 \over 2kq} \right )
Q_{j \pm {1 \over 2}} \left ( {p^2 + q^2 \over 2pq} \right )
\end{array} \right ]
\right \}.
\end{equation}

.

The set of two equations with the top signs is transformed into the set of two
equations with the bottom signs and vice versa with the following replacements:
\setcounter{equation}{21}
\begin{equation}
G^{(+)}\leftrightarrow
F^{(-)},\hspace{.5in}F^{(+)}\leftrightarrow
-G^{(-)},\hspace{.5in}\epsilon\leftrightarrow -\epsilon
\end{equation}

\noindent As a consequence only the equations for $F^{(+)} $and $G^{(+)} $ need be
solved.  For notational convenience the superscripts on $F^{(+)}$ and $G^{(+)}$ are
omitted in future equations.

~

\setcounter{equation}{0}
\setcounter{chapter}{3}
\noindent\begin{Large}{\bf 3 Solutions to the Nonrelativistic Reduction\\
\makebox[.0cm]\ of the Bethe-Salpeter Equation}
\end{Large}

~

In this section the nonrelativistic reduction of the Bethe-Salpeter equation, which is just the
Schr\"odinger equation, is solved numerically in momentum space.  Although the equation can be solved
analytically, here it is solved numerically to illustrate techniques that will be used to solve the
Bethe-Salpeter equation in the instantaneous approximation, an equation that, apparently, cannot in
general be solved analytically.  A method is introduced for handling the singularity in the kernel that
makes possible the use of basis functions that automatically satisfy the boundary conditions both at
small and large momenta.  For angular momentum states $\ell=0$ and $\ell=1$, these basis functions are
not an improvement over the basis functions used by Spence and Vary [7] that only automatically satisfy 
the boundary conditions for small momentum, but for angular momentum states $\ell>1$, the basis
functions used here converge to a solution much more efficiently.

Once the instantaneous approximation has been made, it is straightforward to make
a nonrelativistic reduction [1].  Keeping the lowest order term in each of the two
interaction terms, proportional to $qQ$ and $(qQ)^2$, respectively,

\begin{eqnarray*}
E^{\prime}\Psi({\bf p}) = ({{\bf p^2} \over {2M}} + {{\bf p^2} \over {2m}})\Psi({\bf
p}) + {{qQ} \over {8\pi^3}}\int_{- \infty}^{\infty}{d^3q \over ({\bf
p}-{\bf q})^2}\hspace{.2cm}\Psi({\bf q})\hspace{2.1cm}
\end{eqnarray*}
\vspace{-.6cm}
\begin{eqnarray*}
\hspace{1.6cm}+(1-8)\hspace{.2cm}{(qQ)^2 \over 2^8\pi^6M}\int_{-
\infty}^{\infty}{d^3q
\over ({\bf p}-{\bf q})^2}\hspace{.2cm}\int_{- \infty}^{\infty}{d^3k \over ({\bf
q}-{\bf k})^2}\hspace{.2cm}\Psi({\bf k}).\hspace{1.7cm}(3.1)
\end{eqnarray*}

\noindent The nonrelativistic energy $E'$ is related to the relativistic energy $E$
by
$E=m+M+E'$.  In the final term in the above equation, the integer ``1''in the first
parenthesis is from single-photon exchange while the integer ``-8'' is from the
``seagull'' interaction.

Fourier transforming (3.1), 
\begin{eqnarray*}
E^{\prime}\Psi({\bf x}) = -({1 \over 2M} + {1 \over 2m}){\bf \nabla}^2\Psi({\bf x})
+ {qQ \over 4\pi}\hspace{.2cm} {1 \over |{\bf x}|}\hspace{.2cm}\Psi({\bf x})
\end{eqnarray*}
\vspace{-0.3cm}
\begin{eqnarray*}
\hspace{3.6cm}+(1-8)({qQ \over 4\pi})^2\hspace{.2cm}{1 \over 4M}\hspace{.2cm}{1 \over
|{\bf x}|^2}\hspace{.2cm}\Psi({\bf x}).\hspace{3.4cm}(3.2)
\end{eqnarray*}
From (3.2) it follows that single photon exchange yields a repulsive potential term
proportional to $(qQ)^2$ that decreases as the square of the distance between the
constituents while the ``seagull'' interaction yields an attractive potential with
the same form that is eight times as strong.  To better compare the solutions here
with those of Spence and Vary [7], in the nonrelativistic limit the potential term
proportional to $(qQ)^2$ will be neglected.  However, the effects of this term will
be significant when (2.13), the Bethe-Salpeter equation in the instantaneous
approximation, is solved in the next section.

To solve (3.1) a dimensionless momentum $\bf p'$ is defined by 
\setcounter{equation}{2}
\begin{equation}
{\bf p'} \equiv {{\bf p} \over \sqrt {-2\mu E'}},
\end{equation}

\noindent where $\mu$ is the reduced mass, $\mu=Mm/(M+m)$.  Omitting the term
proportional to
$(qQ)^2$, (3.1) becomes
\begin{equation} 
\hspace{1.5cm}(1 + {\bf p}^{\prime 2})\hspace{.2cm}\Psi({\bf p}^{\prime}) = {qQ \over
4\pi}
{\sqrt {-2\mu E^{\prime}} \over 2\pi^2 E'}\int_{-\infty}^{\infty}{d^3q'
\over ({\bf p'}-{\bf q'})^2}\hspace{.2cm}\Psi({\bf q'}).
\end{equation}

\noindent Eq. (3.4) is the integral form of the Schr\"odinger equation for a
quanta with mass $\mu$ and charge $q$ interacting with a stationary charge $Q$ via
the Coulomb potential.  Since the momentum variables are all now dimensionless, for
notational convenience the primes will be omitted in future equations.  

The solution is of the form
\begin{equation}
\Psi({\bf p})=R(p)Y_\ell(\theta,\phi).
\end{equation}

\noindent Using Hecke's theorem [18,19,20,10] the angular integration is easily performed.  The
angular dependence separates, yielding an integral equation,

\begin{equation}
\hspace{1.5cm}(1 + p^2)pR(p) = {qQ \over 4\pi} \hspace{.2cm}{\sqrt {-2\mu E'} \over
\pi E'}
\int_{0}^{\infty}dq\hspace{.2cm}Q_{\ell}({p^2 + q^2 \over
2pq})\hspace{.2cm}qR(q). 
\end{equation}
\noindent Defining
\begin{equation}
\lambda \equiv {qQ \over 4\pi}\hspace{.2cm}{\sqrt {{-\mu} \over
2E'}},
\end{equation}

\noindent(3.6) becomes
\begin{equation} 
(1 + p^2)pR(p) = \lambda{2 \over \pi}\int_{0}^{\infty}
dq\hspace{.2cm}Q_{\ell}({{p^2 + q^2} \over
{2pq}})\hspace{.2cm}qR(q).
\end{equation}

\noindent Theoretically,
\begin{equation}
\lambda=\ell+n; \hspace{.5in}n=1,2,\dots .  
\end{equation}
Eq. (3.8) would be straightforward to solve numerically were it not for the fact
that $Q_{\ell}((p^2+q^2)/2pq)$ has a logarithmic singularity at $p=q$. 

The boundary conditions are determined with the aid of the asymptotic
relationship for Legendre functions of the second kind [21],
\begin{equation}
Q_{\ell}(z) _{\stackrel{\displaystyle\longrightarrow}  {z \rightarrow
\infty}} {\sqrt {\pi}\hspace{.1cm}\Gamma(\ell + 1) \over 2^{\ell + 1} \Gamma (\ell +
{3 \over 2})}\hspace{.2cm} {1 \over z^{\ell + 1}}.
\end{equation}

\noindent At small $p$ the function $pR(p)$ is assumed to be of the form

\begin{equation}
pR(p) _{\stackrel{\displaystyle\longrightarrow}  {p \rightarrow 0}}
p^{c_0}
\end{equation}

\noindent where $c_0$ is a constant.  From (3.10) it follows
that at small $p$, $Q_{\ell}((p^2+q^2)/2pq) \rightarrow p^{\ell+1}$.  Equating the
left- and right-hand sides of (3.7), at small $p$ the equality $pR(p) \sim
p^{\ell+1}$ is obtained, implying 
\begin{equation}
c_0=\ell+1.
\end{equation}

\noindent At large $p$ the function $pR(p)$ is assumed to be of the form

\begin{equation}
pR(p) _{\stackrel{\displaystyle\longrightarrow}  {p \rightarrow
\infty}} {1
\over p^{c_\infty}}.
\end{equation}

\noindent Using logic analogous to that which lead to (3.12), 
\begin{equation}
c_{\infty}=\ell+3.
\end{equation}
Solutions are obtained by expanding $pR(p)$ in terms of $N$ cubic B-splines $B_j(p)$
[8],
\begin{equation}
pR(p)=F(p)\sum_{j=1}^N c_jB_j(p).
\end{equation}
By choosing the convergence function $F(p)$ in (3.15) so that  at small and large $p$ it behaves as the
solution $pR(p)$ itself, fewer B-splines are required to represent solutions that go to zero
rapidly at the boundaries. 

Cubic B-splines are defined on five
consecutive knots.  To determine the spacing of the knots, $N-4$ zeros $x_i$ of a Chebychev polynomial
are calculated using the formula
\begin{equation}
x_i=-cos{(2i-1)\pi \over 2(N-4)},\hspace{.5in} i = 1, 2, \dots, N-4,
\end{equation}

\noindent and then the knots $T_{i+4}$ on the positive $p$-axis are determined by

\begin{equation}
T_{i + 4} = C_1 \sqrt {{1 + x_i \over 1 - x_i}} +C_2,
\hspace{.5in} i = 1, 2, \dots, N-4,
\end{equation}

\noindent where $C_1$ and $C_2$ are constants.  The knot $T_4$ is placed at the
origin and three knots are placed on the ``negative'' $p$-axis to create maximum
freedom in constructing the solution $pR(p)$ near the origin.  The three knots on
the ``negative'' $p$-axis are mirror images of the first three knots in (3.17).  

Spence and Vary [7] note that integrals of the form
\begin{eqnarray*}
\hspace{2.5cm}\int_{0}^{\infty} dp\hspace{.1cm} p^{\ell+k}Q_\ell({p^2+q^2 \over
2pq}),\hspace{.5in}k=0,1,2,...\hspace{2.2cm}(3.18)
\end{eqnarray*}
\noindent are both finite and readily calculated analytically, so they choose $F(p)$
in (3.15) to be given by $F(p)=p^{\ell+1}$, which, from (3.12), automatically
satisfies the boundary condition near $p=0$ provided that the sum of B-splines in
(3.15) are non-zero and slowly changing near the origin.  Since the  B-splines 
vanish at the largest knot, an appropriate sum of B-splines will satisfy the boundary
condition at large
$p$.  However, the boundary conditions are satisfied both at small and at large $p$
with the choice
\setcounter{equation}{18}
\begin{equation}
F(p)={p^{\ell+1} \over (c^2+p^2)^{\ell+{3 \over
2}}}
\end{equation}
\noindent where $c$ is a constant.  Note that at small $p$, $F(p) \rightarrow p^{\ell+1}$ as expected
from (3.12), while at large $p$,
$F(p)\rightarrow p^{-(\ell+2)}$, which is one power of
$p$ less than is indicated in (3.14).  Since the last B-spline in the expansion (3.15) vanishes at the
largest knot, the boundary conditions will be satisfied automatically both for small and large momenta
provided that the sum of B-splines in (3.15) is slowly changing at small momenta and goes to zero as the
reciprocal of the momentum at large momenta.  By choosing
$F(p)$ in (3.15) appropriately, solutions that decrease rapidly at small and large momenta can be
accurately reproduced with fewer B-splines than when  $F(p)$ is omitted.

Eq. (3.8) is solved numerically by converting the integral eigenvalue equation
into a generalized matrix equation using the Rayleigh-Ritz-Galerkin method [21]. The
solution is expanded in terms of B-splines using (3.15), and then both sides of
(3.8) are multiplied by
$B_i(p)F(p)$ and integrated over $p$.  A generalized matrix equation results that is
of the form $Ac=\lambda Bc$ where the matrices $A$ and $B$ are given,
respectively, by    
\begin{eqnarray*}
\hspace{2.4cm}A_{ij}=\int_{0}^{\infty} dp B_i(p)F(p)(1+p^2)F(p)B_j(p)
\hspace{2.4cm}(3.20a)
\end{eqnarray*}
and 
\begin{eqnarray*}
\hspace{1.2cm}B_{ij}={2 \over \pi}\int_{0}^{\infty} dpB_i(p)F(p)\int_{0}^{\infty} dq
Q_\ell({p^2+q^2 \over 2pq})F(q)B_j(q). \hspace{1.2cm}(3.20b)
\end{eqnarray*}
The elements of the column vector $c$ are the expansion coefficients $c_j$ in
(3.15). Since both of the above matrices are symmetric and $A_{ij}$ is positive
definite, the eigenvalues are real [16].

The choice $F(p)=p^{\ell+1}$ works well for small values of angular momentum because
the sum of a small number of B-splines readily creates a function that 
decreases as $p^{-(2\ell+4)}$ at large $p$, thus satisfying the boundary condition
as given in  (3.14).  In addition, when $F(p)=p^{\ell+1}$, the integrals
over $Q_{\ell}((p^2+q^2)/2pq)$ in (3.20$b$) can be performed analytically because they
are of the form (3.18).  For larger values of angular momentum, however, the
choice $F(p)=p^{\ell+1}$ does not work well because the sum of a small number of
B-splines does not readily create a function that decreases sufficiently rapidly at
large momentum so as to satisfy the boundary conditions.

The choice (3.19) for the convergence function immediately allows the boundary conditions to be
satisfied by a sum of B-splines that is slowly changing, but now the integrals over $q$ in (3.20$b$) can
no longer be  readily performed analytically.  The method for integrating the variable $q$ over the
singularity in the integrals is simple conceptually but somewhat involved numerically:  Except in an
$\epsilon$-neighborhood of the singularity, all integrations are performed numerically using Gaussian
quadrature with a seven-point option.  As the integration variable approaches the singularity, where the
integrand changes most rapidly, integration intervals are decreased to maintain accuracy of the
numerical integration. Within an $\epsilon$-neighborhood of the singularity, the integrand, excluding the
associated Legendre function
$Q_{\ell}((p^2+q^2)/2pq)$, is expanded in a Taylor series about the singularity. The integral is then a
sum of integrals of the form (3.18) that can be integrated analytically.  The parameter $\epsilon$ is
chosen to be the smaller of 0.01 or the distance from the singularity to the nearest knot, thus avoiding
the complication of integrating over a knot.  

To obtain a numerical estimate of the accuracy of each solution, the left- and
right-hand sides of (3.6) are calculated midway between each pair of knots on the
(positive) $p$-axis.  A reliability coefficient $R$ [14], which is a statistical measure of
how closely the two sides of the equation agree at the selected points, is
calculated.  If the two sides of the equation agree exactly at all of the
selected points, then $R$ equals unity.  Determining where the left- and right-hand sides of the
equation agree least well reveals possible problems with trial solutions.

When  $F(p)=p^{\ell+1}$, excellent solutions are obtained for $\ell=0$ and
$\ell=1$.  With 21 splines in the expansion (3.15), eigenvalues were obtained with
four or five significant figures, and corresponding $R$-values were in the range
$0.999<R<1.00$.  However, when $\ell=2$ as shown in second and third columns of Table
3.1, an incorrect eigenvalue appears with a corresponding $R=0.00112$.  For
$\ell=3$, the first few eigenvalues are accurate, but the corresponding $R$-values are
on the order of $10^{-3}$, indicating the solutions are unreliable.  Examination of
these solutions reveals that they are incorrect at large $p$.  By choosing the
convergence factor $F(p)$ as given in (3.18), the difficulties that appeared for 
$\ell > 1$ are eliminated as can be seen from the fourth and fifth columns of Table
3.1.      

~

\noindent Table 3.1:  Numerical values of $\lambda= \ell+1, \ell+2,...$ when 21
splines are used in the expansion (3.15).
 
~

\hspace{1.7cm}\begin{tabular}{|c|c|c|c|c|}\hline
\multicolumn{1}{|}{|c|}&\multicolumn{2}{|c|}{$F(p)=p^{\ell+1}$}
&\multicolumn{2}{|c|}{$F(p)={{p^{\ell+1}} /over {(c^2+p^2)^{\ell+3/2}}}$}\\
\hline
$\ell$  & $\lambda$ & R & $\lambda$ & R\\
\hline
\hline
\ 2 & 3.00000&0.99884& 3.00000&1.0000\\
\hline
\ &4.00007&0.97013&4.00001&1.0000\\ 
\hline
\  &4.49547&0.00112&5.00003&1.0000\\
\hline
\  &5.00030&0.84320&6.00008&1.0000\\
\hline
\hline
\ 3&4.00118&0.00266&4.00000&1.0000\\
\hline
\  &5.00844&-0.00028&5.00000&1.0000\\
\hline
\  &6.02762&-0.00031&6.00004&1.0000\\
\hline
\  &7.06566&-0.00023&6.99999&1.0000\\
\hline
\end{tabular}

\vspace {.25in}

~

\setcounter{chapter}{4}
\noindent{\Large {\bf 4 Solutions of the Bethe-Salpeter Equation \\ 
\ in the Instantaneous Approximation }}

~

Solutions to the Bethe-Salpeter equation in the instantaneous approximation are obtained using two
different basis systems.  The first basis system is comprised of essentially the same basis functions
that were employed to calculate solutions to the  nonrelativistic hydrogen atom. Because the basis
functions vanish at large momenta, they are particularly suitable for representing solutions that have 
significant support only to moderately large values of momentum (and position).  For this basis system
four B-splines are non-zero between consecutive knots in the physical region except for the final four
knots at the largest values of momentum:  There the number of non-zero B-splines between consecutive
knots decreases from three to two until only one B-spline is non-zero between the final two knots, thus
making it increasingly difficult to express a solution at large momentum in terms of these basis
functions.  To better represent solutions that are highly localized in position space and, therefore,
have significant support at large values of momentum, a second basis system is used in which some
basis functions  vanish only at infinite values of momentum and  four  B-splines are non-zero
between consecutive knots in the physical region.

The boundary conditions as $p$ approaches zero and infinity are determined using the
same procedure employed in the previous section.  The results are as follows:  
\begin{eqnarray*}
\hspace{1.0in}pG(p) _{\stackrel{\displaystyle\longrightarrow}  {p \rightarrow
0}} p^{j+{1 \over2}}
\hspace{1.0in}
pF(p) _{\stackrel{\displaystyle\longrightarrow}  {p \rightarrow 0}} p^{j+{3 \over2}}
\hspace{.9in} (4.1a)
\end{eqnarray*}
\begin{eqnarray*}
\hspace{1.0in}pG(p) _{\stackrel{\displaystyle\longrightarrow}  {p \rightarrow
\infty}} {1 \over p^{j+{3 \over 2}}}
\hspace{1.0in}
pF(p) _{\stackrel{\displaystyle\longrightarrow}  {p \rightarrow \infty}} {1
\over p^{j+{5 \over 2}}}
\hspace{.8in} (4.1b)
\end{eqnarray*}
\noindent Solutions can be obtained using methods of the previous section and are of the form 
\begin{eqnarray*}
\hspace{1.2in}pG(p)={\cal G}_1(p)\sum_{j=1}^Ng_jB_j(p),\hspace{4.7cm}(4.2a) 
\end{eqnarray*}
\noindent and
\begin{eqnarray*}
\hspace{1.2in}pG(p)={\cal F}_1(p)\sum_{j=1}^Nf_jB_j(p).\hspace{4.7cm}(4.2b)
\end{eqnarray*}
The convergence functions ${\cal G}_1(p)$ and ${\cal F}_1(p)$ are chosen so that the
boundary conditions are automatically satisfied provided that the sums of B-splines in
the previous equation are slowly changing for small and large momenta. 
\setcounter{equation}{2}
\begin{equation}
{\cal G}_1(p)={p^{j+{1 \over 2}} \over (c_{\cal G}^2+p^2)^{j+{1 \over 2}}} \hspace{.5in} {\cal
F}_1(p)={p^{j+{3 \over 2}}
\over (c_{\cal F}^2+p^2)^{j+{3 \over 2}}}
\end{equation}
\noindent In the above equation $c_{\cal G}$ and $c_{\cal F}$ are constants. Note that at small $p$, 
$p{\cal G}_1(p)$ and $p{\cal F}_1(p)$ vanish as indicated in (4.1$a$), but at large $p$ they decrease by
a factor $p$ more slowly than indicated in (4.1$b$) because the B-splines themselves vanish at large
$p$.  As can be seen from (4.3), solutions go to zero rapidly at the boundaries even at the smallest
value
$j=1/2$, so it would be difficult to obtain any accurate solutions in the instantaneous approximation
without using convergence functions. 

Eq. (2.21) is converted into a generalized matrix equation of the form
\begin{eqnarray*}
\hspace{4.4cm} A\left [\begin{array}{c}g\\f\end{array}\right]
&=&\epsilon~ B\left [\begin{array}{c}g\\f\end{array}\right]
\hspace{4.0cm}(4.4)\\ 
\end{eqnarray*}
\noindent by multiplying the top and bottom equations by ${\cal G}_1(p)B_i(p)$ and
${\cal F}_1(p)B_i(p)$, respectively, and then integrating over $p$.  The elements of
the column vectors $g$ and $f$ are, respectively, the expansion coefficients
$g_j$ and $f_j$ in (4.2).  Since the matrices $A$ and $B$ have been constructed so
that both are symmetric and $B$ is positive definite, the dimensionless energy
eigenvalue $\epsilon$ is forced to be real [16] as required.

The Bethe-Salpeter equation in the instantaneous approximation contains double integrals while in the
nonrelativistic limit, the equation involves only single integrals. In spite of this complication, by
performing integrations in a specific order, all integrals with  a logarithmic singularity that are
necessary to solve the equation are of the form already encountered in the previous section.
However, the technique used to integrate over the logarithmic singularity in the previous section fails
at very large values of momentum:  All of the terms being
expanded in a Taylor series about the singular point $p$ are functions of $q^2$ with the result that a
typical term in the expansion is $a_n(q^2-p^2)^n$.  Within an $\epsilon$-neighborhood of $p$, the
maximum value of  
 $q^2-p^2$ is $2p\epsilon$.  When $p$ is of the order of $1/ \epsilon$ the expansion begins to lose
accuracy.  By choosing $ \epsilon$  to decrease linearly with increasing $p$, this problem is avoided.
 
To check the accuracy of the solutions by calculating the reliability
coefficient, double integrals of the following form must be evaluated:
\setcounter{equation}{4}
\begin{equation}
\int_{0}^{\infty} {dq \over {\omega_-(q)}} \hspace{.2cm} Q_{\ell}({p^2 + q^2
\over 2pq})
\int_{0}^{\infty}
dk \hspace{.2cm} Q_{\ell}({k^2 + q^2) \over
2kq}) {q^{\ell+d_1} \over {(c^2+k^2)^{d_2}}}B_j(k)
\end{equation}
The integral over the variable $k$ can be calculated as previously discussed.  Except
within an $\epsilon$-neighborhood of the logarithmic singularity of the integrand,
the integral over the variable $q$ is evaluated numerically. Within the $\epsilon$-neighborhood, the
integral over $k$ is expressed as a power series expansion in the variable $q$,   
\begin{equation}
\int_{0}^{\infty}
dk \hspace{.2cm} Q_{\ell}({k^2 + q^2) \over
2kq}) {q^{\ell+d_1} \over
{(c^2+k^2)^{d_2}}}B_j(k)=q^{\ell+1}\sum_{j=0}^3a_j(q-p)^j.
\end{equation}
The power series expansion in the above equation depends on the fact that the integral
vanishes as $q^{\ell+1}$ at small $q$ , a fact that is readily verified using
(3.10).  The coefficients
$a_j$ are determined numerically so that the expansion and the integral agree
at $p+\epsilon$, $p+\epsilon/3$, $p-\epsilon/3$ and $p-\epsilon$.  Using the
expansion in (4.6), within the
$\epsilon$-neighborhood of the logarithmic singularity at $p$, the integral (4.5)
can be performed analytically.

To better represent solutions that are highly localized in position space and, therefore, have
significant support at large values of momentum, a second basis system is introduced in which some basis
functions  vanish only at infinity.  To construct the basis system, the momentum is first mapped onto a
compact space with the transformation 
\setcounter{equation}{6}
\begin{equation}
x(p) =b\hspace{.1cm}{p^2-a \over p^2+a }, 
\end{equation}
\noindent where $a$ and $b$ are constants. From (4.7) it follows that $-b\leq x(p) \leq b$.

The knots are determined by first calculating $N-8$ zeros $x_i$ of a Chebychev polynomial using the
formula
\begin{equation} x_i=-cos{(2i-1)\pi \over 2(N-8)},\hspace{.5in} i = 1, 2, \dots, N-8.
\end{equation}
\noindent The knots in the region $-b$ $<$ x $<$ $b$ are then given by
\begin{equation}
T_{i + 4} = bx_i,\hspace{.5in} i = 1, 2, \dots, N-8.
\end{equation}
 \noindent The knot $T_4$ is placed at $x=-b$ ($p=0)$ and three knots are placed in the region $x$ $<$
$-b$ (on the ``negative'' $p$-axis) to create maximum freedom in constructing solutions near the origin. 
The three knots in the region $x$ $<$ $-b$ are mirror images of the first three knots in (4.9).  In a
similar fashion the knot $T_{N-3}$ is placed at $x=b$ ($p=\infty)$ and three knots are placed in the
region $x$ $>$ $b$ (``$p$ $>$ $\infty$'') to create maximum freedom in constructing solutions at very
large momenta. The three knots in the region $x$ $>$ $b$ are mirror images of the final three knots in
(4.9).  With the above knot structure, four B-splines are non-zero between each pair of adjacent knots
in the entire physical region $0\leq p \leq \infty$.   
 
The solution is expanded in terms of B-splines as follows:

\begin{eqnarray*}
\hspace{3.5cm}pG(p)={\cal G}_2(p(x))\sum_{j=1}^Ng_jB_j(x),\hspace{3.5cm}(4.10a) 
\end{eqnarray*}
\noindent and
\begin{eqnarray*}
\hspace{3.5cm}pG(p)={\cal F}_2(p(x))\sum_{j=1}^Nf_jB_j(x),\hspace{3.5cm}(4.10b)
\end{eqnarray*}
\noindent where
\setcounter{equation}{10}
\begin{equation}
{\cal G}_2(p)={p^{j+{1 \over 2}} \over (c_{\cal G}^2+p^2)^{j+1}},  \hspace{.5in} {\cal
F}_2(p)={p^{j+{3 \over 2}}
\over (c_{\cal F}^2+p^2)^{j+2}}.
\end{equation}

\noindent  For the second basis system the final three B-splines in the expansion are non-zero at $x=b$
($p=\infty$).  Consequently, solutions that have significant support at large values of momentum
are more readily expressed in terms of the second set of basis functions.  Since some B-splines are
finite at
$p=\infty$, the functions ${\cal G}_2(p)$ and ${\cal F}_2(p)$ are chosen to satisfy the boundary
conditions (4.1) both at small and large momenta.

To integrate over singularities at large values of momentum $p$, the integrand is
expanded in a Maclaurin series in the variable $1/q$ instead of in a Taylor series.  Specifically, if
the location of the first knot less than the singularity corresponds to a value of momentum equal to
or less than 50, then integrals are evaluated as previously discussed. On the other hand, if the
first knot less than the singularity corresponds to a value of momentum greater than 50, the integral is
evaluated numerically from $x=-b$ ($p=0$) to the knot.  From the knot to $x=b$ ($p=\infty$), the
integral is evaluated analytically by expanding the integrand, excluding the Legendre
function of the second kind, in a Maclaurin series. The necessary formulas for carrying out the
integration are given in the appendix.

A corresponding modification is required to evaluate the double integrals (4.5).  When the location of
first knot less than the logarithmic singularity at $q=p$ corresponds to a value of $p$ equal to or
less than 50, the integral is evaluated as before.  When
the position of the knot corresponds to a value of $p$ greater than than 50, the integral is evaluated
numerically except within an $\epsilon$-neighborhood of the singularity by expanding the integral over
$k$ as a Maclaurin series in the variable $1/q$,   
\begin{equation}
\int_{0}^{\infty}
dk \hspace{.2cm} Q_{\ell}({k^2 + q^2) \over
2kq}) {q^{\ell+d_1} \over
{(c^2+k^2)^{d_2}}}B_j(k)={1 \over q^{\ell+1}}\sum_{j=0}^3a_j{1 \over q^j}.
\end{equation}
In writing the expansion,  the fact has been used that the
integral vanishes as ${1/q^{\ell+1}}$ at large $q$.  The coefficients
$a_j$ are determined numerically so that the expansion and the integral agree
at $p+\epsilon$, $p+\epsilon/3$, $p-\epsilon/3$ and $p-\epsilon$.  Using the
expansion in (4.12), within the
$\epsilon$-neighborhood of the logarithmic singularity at $p$, the integral (4.5)
can be performed analytically.

The first basis system has more knots concentrated at small values of momentum and, therefore, is more
suitable for representing weakly-bound solutions or solutions with detailed structure in this
region.  The second basis system has more knots at large values of momentum and is better for
representing strongly-bound solutions that have significant support at large values of momentum. 
However, at least when 35 or fewer splines are used, the second basis system does not adequately
represent the most strongly bound solutions with $\epsilon$  on the order of or less than about $0.3$. 
Of course, for such states the instantaneous approximation is not a satisfactory approximation to the
Bethe-Salpeter equation.   

When only the effects of single-photon exchange are included, the solutions in the instantaneous
approximation are approximate solutions of the Bethe-Salpeter equation in the ladder approximation.  If
the masses of the constituents are equal, zero-energy, analytical solutions exist for the Bethe-Salpeter
equation in the ladder approximation for several values of $qQ/4\pi$ in the range $0 < qQ/4\pi <
100$ [10].  In the instantaneous approximation, when only single-photon exchange is included, no
bound-state solutions were found with either zero or finite energy within this range of coupling
constants.

For all data graphed below, the constituent masses are equal although solutions with unequal constituent
masses are no more difficult to determine.  Solutions were calculated using the two different basis
systems previously discussed.  For values of $\epsilon= E/(M+m) > 0.95$, the graphed results are those
obtained from the first basis system, which has more knots at small momenta. For all other values of
$\epsilon$, the graphs are an average of the solutions obtained from the two basis systems. Solutions for
$\epsilon$ obtained from the two different basis systems almost always agreed within 0.04 and usually
agreed more closely while reliability coefficients were almost always greater than
$0.99$.  Solutions were calculated by expanding the wave function in
terms of 35 B-splines.

\setcounter{figure}{0}
\begin{figure}
\begin{tabular}{cccc}
\epsfig{file=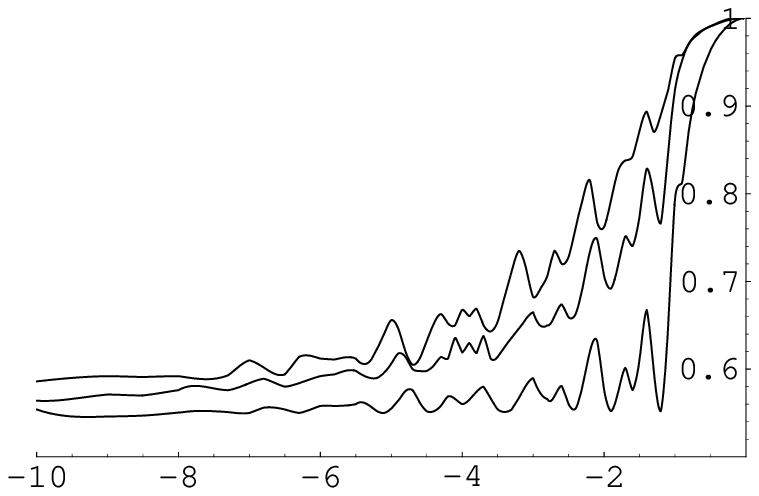,width=6.2cm} &&\epsfig{file=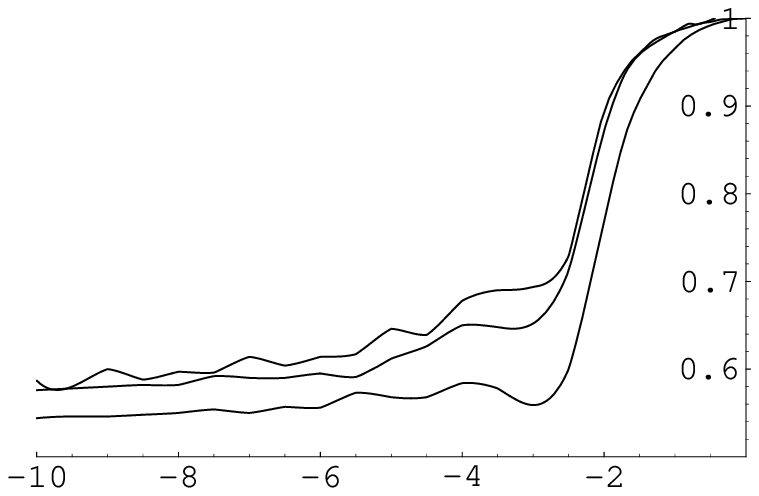,width=6.2cm}& \\
\vspace{-3.0cm}&&&\\
&\hspace{-.5cm}$\epsilon$&&\hspace{-.5cm}$\epsilon$ \\
\vspace{1.5cm}&&&\\
{$qQ/4 \pi$} & &{$qQ/4 \pi$}&\\
(a)& & (b)&\\
\end{tabular}
\caption {The dimensionless energy eigenvalues $\epsilon = E/(M+m)$  of the three lowest
bound states as a function of the coupling constant $qQ/4\pi$ when only single-photon exchange is
included in the instantaneous approximation.  Graphs (a) and (b) correspond to angular momentum
$j=1/2$ and $j=3/2$, respectively.
\label{fig}}
\end{figure}

From Fig. 4.1, as the coupling constant decreases in magnitude, the repulsive effects of angular
momentum become apparent  so that states with higher angular momentum are more weakly bound.

\setcounter{figure}{1}
\begin{figure}
\begin{tabular}{cccc}
\epsfig{file=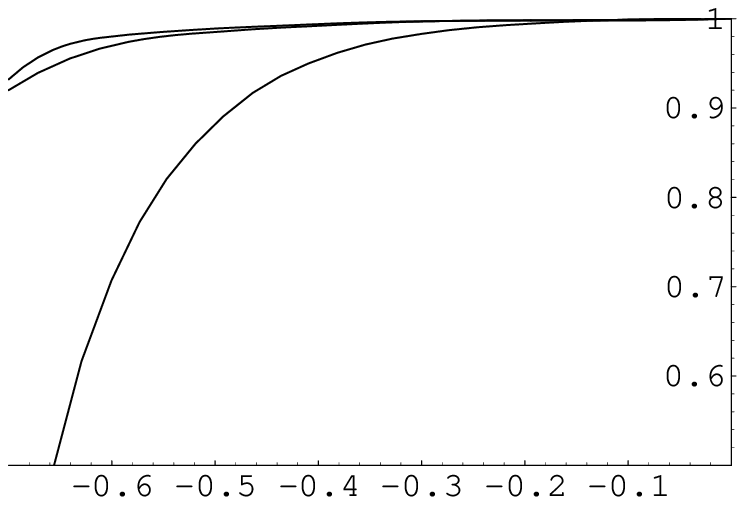,width=6.0cm} &&\epsfig{file=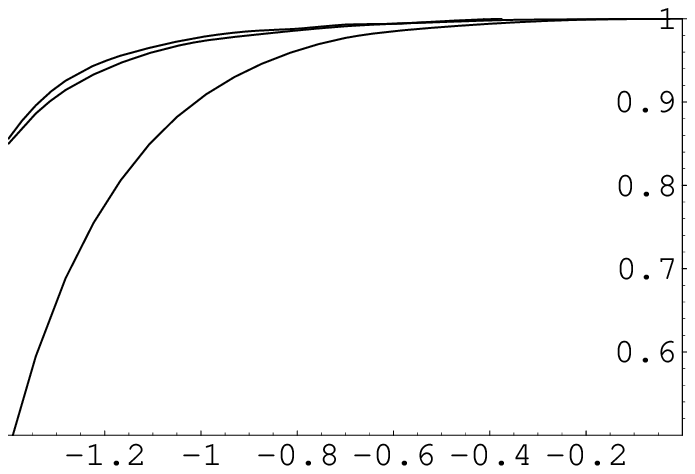,width=6.0cm}& \\
\vspace{-3.0cm}&&&\\
&\hspace{-.7cm}$\epsilon$&&\hspace{-.5cm}$\epsilon$ \\
\vspace{1.5cm}&&&\\
{$qQ/4 \pi$} & &{$qQ/4 \pi$}\\
&(a) & (b)&\\
\end{tabular}
\caption {The dimensionless energy eigenvalues $\epsilon = E/(M+m)$  of the three lowest
bound states as a function of the coupling constant $qQ/4\pi$ in the instantaneous approximation. 
Graphs (a) and (b) correspond to angular momentum $j=1/2$ and $j=3/2$, respectively.
\label{fig}}
\end{figure} 

Because the bound states in Fig. 4.2 include the ``seagull'' term, for the same value of the coupling
constant these states are significantly more tightly bound than the corresponding states in Fig. 4.1,
which only include single-photon exchange.

\begin{figure}
\begin{tabular}{cccc}
&\hspace{-.3cm}\epsfig{file=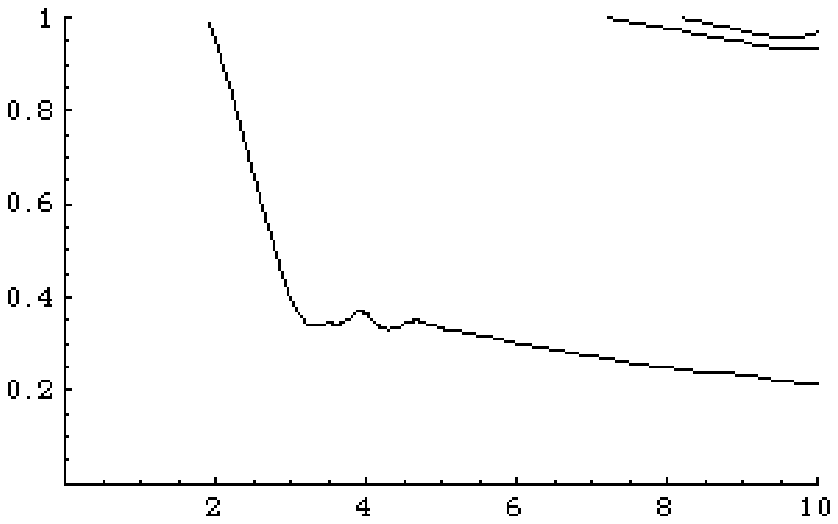,width=5.6cm} &&\hspace{-.3cm}\epsfig{file=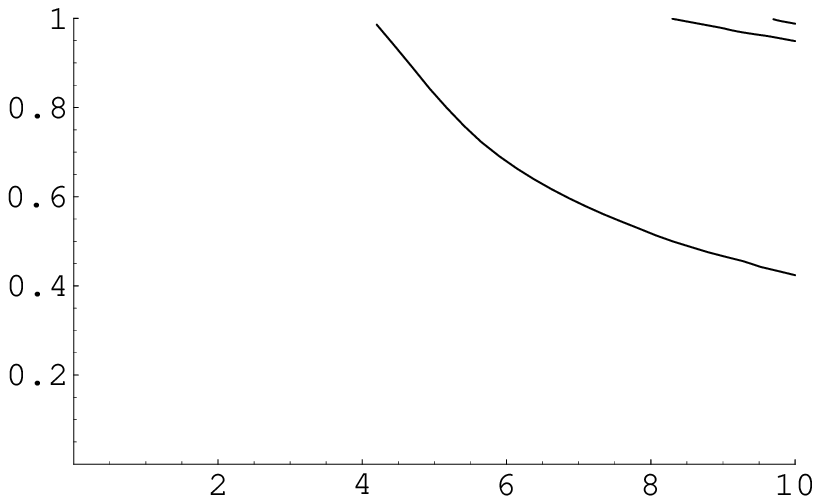,width=5.6cm} \\
\vspace{-3.0cm}&&&\\
$\epsilon$&&$\epsilon$&\\
\vspace{1.5cm}&&&\\
&{$qQ/4 \pi$}  &{$qQ/4 \pi$}&\\
&(a)&  (b)&\\
\end{tabular}

\caption {The dimensionless energy eigenvalues $\epsilon = E/(M+m)$  of the three lowest
bound states as a function of the coupling constant $qQ/4\pi$ in the instantaneous approximation. 
Graphs (a) and (b) correspond to angular momentum $j=1/2$ and $j=3/2$, respectively.
\label{fig}}
\end{figure} 
 
The bound states in Fig. 4.3 that occur for positive values of $qQ/4\pi$ are the result of the 
``seagull'' interaction, which is always attractive nonrelativistically.  As a consequence, like-signed
electric charges can attract [10]. But, at least for equal-mass constituents in the instantaneous
approximation as shown in Fig. 4.3(a), no binding occurs between like-signed charges for values of
$qQ/4\pi < 1.9$.  Since no ``elementary'' constituent could have such a large charge, the effect would
not be observable.  On the other hand, by considering the fully relativistic equation with unequal-mass
constituents, if such binding could be achieved for constituents that have charges with magnitudes on
the order of $e$, the effect might be observable.

~ 

\setcounter{chapter}{5}
\noindent\begin{Large}{\bf 5 Conclusions}
\end{Large}

~

In spite of a derivative coupling in the Lagrangian, the Bethe-Salpeter equation, which includes the
effects of both single-photon exchange and the ``seagull'' interaction, can readily be solved in the
instantaneous approximation: Terms appear symmetrically in the equation so that it can be converted to a
generalized matrix eigenvalue equation of the form $Ac=\epsilon Bc$ where the matrices $A$ and $B$ are
symmetric and $B$ is positive definite, a sufficient condition for yielding real eigenvalues $\epsilon$. 
Using a generalization of a method introduced by Spence and Vary [7] for handling logarithmic
singularities in integrands, basis systems are used that automatically satisfy the boundary conditions,
making it possible to obtain solutions in the instantaneous approximation.  Weakly-bound solutions are
more efficiently obtained by writing the solution in terms of basis functions that extend only to finite
values of momentum while solutions with large binding energies are obtained more accurately by
mapping the momentum onto a compact space and then expressing the solutions in terms of basis functions,
some of which vanish only at infinity.  A statistical measure is used to provide an indication of the
accuracy of solutions by determining how well the wave functions satisfy the equation midway between
each knot in the physical region.
\bigskip

\noindent{\bf Acknowledgments}

\medskip

I would like to thank Professor John J. Skowronski for very helpful discussions
regarding statistical parameters that would indicate the reliability of
numerical solutions to equations. Dr. David G. Robertson assisted in optimizing the code.  This
work was supported by a grant of computer time from the Ohio Supercomputer Center.

~

\bigskip
\noindent\begin{Large}{\bf Appendix: Calculation of Integrals}
\end{Large}

~
 
Here the formulas are given for calculating the integrals of the form
\begin{eqnarray*}
\hspace{.7cm}B^{\hspace{.11cm}\ell,k}_{(a,b)}(p)\equiv\int_a^bdqQ_\ell\hspace{.05cm}({p^2+q^2 \over
2pq}){1
\over q^{\ell+k}}\hspace{1.0cm} \begin{array}{c}
\ell=0;\hspace{.3cm} k=1,2,3,\dots\\  \ell\ge1;\hspace{.3cm}k=0,1,2,\dots\end{array}
\hspace{.7cm}(A1) 
\end{eqnarray*}
that are required to integrate over the logarithmic singularities when they occur at large $p$.  Using
the recursion formula for Legendre functions of the second kind [23],
\begin{eqnarray*}
\hspace{2.7cm}Q_{\ell+1}(z) = {2\ell+1 \over \ell+1}zQ_\ell(z)-{\ell \over
\ell+1}Q_{\ell+1}(z),\hspace{2.7cm}(A2) 
\end{eqnarray*}
a recursion relation for $B^{\hspace{.11cm}\ell,k}_{(a,b)}(p)$ follows immediately:
\begin{eqnarray*}
\hspace{.5cm}B^{\hspace{.11cm}\ell+1,k}_{\hspace{.11cm}(a,b)}(p)={2\ell+1 \over \ell+1}
[{p \over2}B^{\hspace{.11cm}\ell,k+2}_{\hspace{.11cm}(a,b)}(p)
+{1 \over2p}B^{\hspace{.11cm}\ell,k}_{(a,b)}(p)]
-{\ell \over \ell+1}B^{\hspace{.11cm}\ell-1,k+2}_{\hspace{.3cm}(a,b)}(p)
\hspace{.5cm}(A3) 
\end{eqnarray*}
The integrals $B^{\hspace{.11cm}\ell,k}_{(a,b)}(p)$ are readily expressed in terms of the
integrals
\begin{eqnarray*}
\hspace{4.1cm}I^{\hspace{.19cm}k}_{(a,b)}(p)\equiv \int_a^bdq\hspace{.1cm}{ln(q+p) \over
q^k}.\hspace{3.9cm}(A4)
\end{eqnarray*}
\noindent Specifically,
\begin{eqnarray*}
\hspace{3.4cm}B^{\hspace{.11cm}0,k}_{(a,b)}(p)=I^{\hspace{.19cm}k}_{(a,b)}(p)-I^{\hspace{.19cm}k}_{(a,b)}(-p)
\hspace{3.4cm}(A5)
\end{eqnarray*}
\noindent and
\begin{eqnarray*}
\hspace{1.0cm}B^{\hspace{.11cm}1,k}_{(a,b)}(p)=
{p \over
2}[I^{\hspace{.05cm}k+2}_{\hspace{.05cm}(a,b)}(p)-I^{\hspace{.05cm}k+2}_{\hspace{.05cm}(a,b)}(-p)] +{1
\over 2p}[I^{\hspace{.19cm}k}_{(a,b)}(p)-I^{\hspace{.19cm}k}_{(a,b)}(-p)]
\hspace{1.0cm}\\
\end{eqnarray*}
\vspace{-1.2cm}
\begin{eqnarray*}
\hspace{2.7cm}+ \left \{\begin{array}{c}-ln({b \over a}) \hspace{1.1cm}if \hspace{.2cm}k=0\\{1 \over
k}[{1 \over b^k}-{1 \over a^k}] \hspace{.6cm}if\hspace{.2cm} k>0\end{array}\right\}.
\hspace{4.3cm}(A6)\\ 
\end{eqnarray*}
The integrals $I^{\hspace{.19cm}0}_{(a,b)}(p)$ and $I^{\hspace{.19cm}1}_{(a,b)}(p)$ are calculated
using standard tables of integrals [23], although $I^{\hspace{.19cm}1}_{(a,b)}(p)$ is evaluated as an
infinite series.  For $k \ge 2$, the integral $I^{\hspace{.19cm}k}_{(a,b)}(p)$ can be calculated using
the following formula:  Let 
\begin{eqnarray*}
\hspace{3.6cm}I\equiv \int dx\hspace{.1cm}{ln(a+bx) \over x^k},\hspace{.5cm}k\ge 2.\hspace{3.6cm}(A7)
\end{eqnarray*}
\noindent Integrating by parts,
\begin{eqnarray*}
I={1 \over k-1}\hspace{.1cm}[-{ln(a+bx) \over x^{k-1}}+b \int{dx \over x^{k-1}(a+bx)}].
\end{eqnarray*}
The integral in the above expression is evaluated using partial fractions, yielding the desired formula:
\begin{eqnarray*}
I={1 \over k-1}\hspace{.1cm}[-{ln(a+bx) \over x^{k-1}}+(-{b \over a})^{k-1}ln(a+bx)
\end{eqnarray*}
\vspace{-.5cm}
\begin{eqnarray*}
\hspace{3.0cm}-(-{b \over a})^{k-1}ln(x)+\sum_{j=2}^{k-1}(-{b \over a})^{k-j}{1 \over
(j-1)x^{j-1}}]\hspace{2.2cm}(A8)
\end{eqnarray*}

\clearpage

\noindent\begin{Large}{\bf References}
\end{Large}

\begin{enumerate}
\item    E.E. Salpeter and H.A. Bethe, Phys. Rev. {\bf 84}, 1232 (1951).
\item    G.B. Mainland, Few Body Systems {\bf 26}, 27 (1999).
\item    R. Blankenbecler and R. Sugar, Phys. Rev. {\bf 142}, 1051 (1966).
\item    E.E. Salpeter, Phys. Rev. {\bf 87}, 328 (1952).
\item    J.L. Gammel and M.T. Menzel, Phys. Rev. {\bf A7}, 858 (1973).
\item    D. Eyre and J.P. Vary, Phys. Rev. {\bf D34}, 3467 (1986).
\item    J.R. Spence and J.P. Vary, Phys. Rev. {\bf D35}, 2191 (1987); Phys. Rev. {\bf C47}, 1282 (1993).
\item    See, for example, C. de Boor,  {\it A Practical Guide to Splines} (Springer-Verlag,
         Berlin-Heidelberg-New York, 1978).
\item   J.C. Pati and A. Salam, Phys. Rev. {\bf D10}, 275 (1974);
        J.C. Pati, A. Salam and J. Strathdee, Phys. Lett. {\bf B59}, 265 (1975);
        O.W. Greenberg and J. Sucher, Phys. Lett. {\bf B99}, 339 (1981);
        W. Buchmuller, R.D. Pecci and T. Yanagida, Phys Lett. {\bf B124}, 67 (1983);
        O.W. Greenberg, R.N. Mohapatra and M. Yasue, Phys. Rev. Lett. {\bf 51}, 1737 (1983);
         Nucl. Phys. {\bf B237}, 189 (1984);
        J.C. Pati, Phys. Rev. {\bf D30}, 1144 (1984) 
        M.A. Luty and R.N. Mohapatra, Phys. Lett. {\bf B 396}, 161 (1997).
\item   G.B. Mainland, J. Math. Phys. {\bf 27}, 1344 (1986).         
\item   S. Godfrey and N. Isugur, Phys. Rev. {\bf D32}, 189 (1985); S. Godfrey,
        Nuovo Cimento {\bf A102}, 1 (1989); W. Lucha, F.F. Sch\"oberl and D. Gromes, Phys. Rep. {\bf
        200}, 127 (1991); A.J. Sommerer, J.R. Spence and J.P. vary, Phys. Rev. {\bf C49}, 513 (1994).
\item   S. Hatakeyama {\it et al}, Phys. Rev. Lett. {\bf 81}, 2016 (1998). 
\item   N. Arkani-Hamed and Y. Grossman, Phys. Lett. {\bf B 459}, 179 (1999). 
\item   B.J. Winer, {\it Statistical Principles in Experimental Design} (McGraw Hill, New York, 1962).  
\item  The notation is that of J.D. Bjorken and S.D. Drell, {\it Relativistic Quantum Fields} (McGraw
       Hill, New York, 1965).  $\hbar$ and c are set to unity.  Repeated Greek indices are summed from
       0-3 and
       repeated Roman indices are summed from 1-3.  Bold variables represent vectors in three-dimensional
        space.
\item   See, for example, F.B. Hildebrand, {\it  Methods of Applied
         Mathematics, 2nd Ed.} (Prentice-Hall, Englewood Cliffs, New Jersey, 1965).
\item    C. Schwartz,  Phys. Rev. {\bf 137}, B717 (1965); M.J. Zuilhof and J.A. Tjon, Phys.
         Rev.  {\bf C22}, 2369 (1980);  G. B. Mainland and J.R. Spence, Few Body Systems {\bf 19},
         109 (1995); T. Nieuwenhuis and J.A. Tjon, Few Body Systems {\bf 21}, 167 (1996).   
\item    E. Hecke, Math. Ann. {\bf 78}, 398 (1918).
\item   V. Fock, Z. Phys.{\bf 98}, 145 (1935).
\item   M. Le\'vy, Proc. R. Soc. London Ser.{\bf A204}, 145 (1950).
\item   P.M. Morse and H. Feshbach, {\it Methods of Theoretical Physics} (McGraw-Hill, New York, 1953).
\item   K.E. Atkinson, {\it A Survey of Numerical Methods for the Solution of Fredholm
         Equations of the Second Kind} (SIAM, Philadelphia, 1976);  L.M. Delves and J. Walsh, 
         {\it Numerical Solution of Integral Equations} (Clarendon Press, Oxford, 1974).
\item  I.S. Gradshteyn and I.M. Ryzhik {\it Tables of Integrals, Series, and Products} (Academic
       Press, New York, 1965).

\end{enumerate}

\end{document}